\documentclass[preprint,pra,aps,superscriptaddress,groupedaddress,showkeys]{revtex4-1}
\usepackage{amsmath,amssymb,bbm,hyperref}
\usepackage{dcolumn}
\usepackage{bm}
\usepackage{amssymb}
\usepackage{graphicx}
\usepackage{amssymb}

\begin{document}

\title{On an electrodynamic origin of quantum fluctuations}

\author{{\'A}lvaro G. L{\'o}pez}
\email{alvaro.lopez@urjc.es}
\affiliation{Nonlinear Dynamics, Chaos and Complex Systems Group\\Departamento de F\'isica, Universidad Rey Juan Carlos, Tulip\'an s/n, 28933 M\'ostoles, Madrid, Spain}

\begin{abstract}
In the present work we use the Li\'enard-Wiechert potential to show that very violent fluctuations are experienced by an electromagnetic charged extended particle when it is perturbed from its rest state. The feedback interaction of Coulombian and radiative fields among the different charged parts of the particle makes uniform motion unstable. As a consequence, we show that radiative fields and radiation reaction produce both dissipative and antidamping effects, leading to self-oscillations. Finally, we derive a series expansion of the self-potential, which in addition to rest and kinetic energy, gives rise to a new contribution that shares features with the quantum potential. The novelty of this potential is that it produces a symmetry breaking of the Lorentz group, triggering the oscillatory motion of the electrodynamic body. We propose that this contribution to self-energy might serve as a bridge between classical electromagnetism and quantum mechanics.
\end{abstract}


\date{\today}

\maketitle

\section{Introduction}

It was shown in the mid-sixties that a dynamical theory of quantum mechanics can be provided based on a process of conservative diffusion \cite{nel66}. The theory of stochastic mechanics is a monumental mathematical achievement that has been carefully and slowly carried out along two decades with the best of the rigors and mathematical intuition \cite{nel85}. However, as far as the authors are concerned, the grandeur of this theoretical effort is that it proposes a kinematic description of the dynamics of quantum particles, based on the theory of stochastic processes \cite{van92}. Just as Bohmian mechanics \cite{boh52,boh522}, it tries to offer a geometrical picture of the trajectory of a quantum particle, which would be so very welcomed by many physicists. In the end, establishing a link between dynamical forces and kinematics is at the core of Newton's revolutionary work \cite{new99}.

Perhaps, the absence of geometrical intuition in this traditional sense, during the development of the quantum mechanical formalism, has hindered the understanding of the underlying physical mechanism that leads to quantum fluctuations. In turn, it has condemned the physicist to a systematic titanic effort of mathematical engineering, designing ever-increasing complicated theoretical frameworks. Despite of providing a very refined explanation of many experimental data, which is the main purpose of any physical theory, needless to say, these frameworks entail a certain degree of obscurantism and a lack of understanding. Concerning comprehension only, quantum mechanics constitutes a paradigm of these kind of paradoxical theories, which imply that the more time that it is dedicated to the their study, the less clear that the physical picture of nature becomes. As it has been pointed out by Bohm, this might be a consequence of renouncing to models in which all physical objects are unambiguously related to mathematical concepts \cite{boh52}.

On the contrary, hydrodynamical experimental models that serve as analogies
to quantum mechanical systems have been developed recently, which allow to clearly visualize how the dynamics of a possible quantum particle might be \cite{cou05,pro06}. These experimental contemporary models share many features with the mechanics of quantum particles \cite{for10,bus13} and, fortunately, they are based on firmly established and understandable principles of nonlinear dynamical oscillatory systems and chaos theory \cite{bus18,ali96}. As it is well accepted, these conceptual frameworks have shaken the grounds of the physical consciousness of many scientists by showing the tremendous complexity of the dynamical motion of rather simple classical mechanical systems, and not so simple as well \cite{ali96}. Doubtlessly, the development of computation has proven to be a fundamental tool in this regard, serving as a microscope to the modern physicist, which allows him to unveil the complex patterns and fractal structures that explain the hidden regularities of chaotic motion \cite{agu09}. Thus, even if we can not experimentally trace a particle's path because we perturb its dynamics by the mere act of looking at it, we can always use our powerful computers to simulate their dynamics.

In the final pages of Nelson's work, it is seductively suggested that a theory of quantum mechanics based on classical fields should not be disregarded, as was originally the purpose of Albert Einstein \cite{nel85}. This aim of providing quantum mechanics with a kinematic description, together with the desire of showing the unjustified belief of electrodynamic fields as a merely dissipative force on sources of charge, and not as an exciting self-force as well, are the two core reasons that have spurred the authors to pursue the present goal. By using a toy model and rather simple mathematics, we show as a main result in what follows that a finite-sized charged accelerated body always carries a vibrating field
with it, what can convert this particle into a stable limit cycle oscillator by virtue
of self-interactions. This implies that the rest state of this charged particle can be unstable, and that stillness (or uniform motion) might not the default state of
matter, but also accelerated oscillatory dynamics. We close this work by deriving an analytical expression of the self-potential. For this purpose we only need to assume that inertia is of purely electromagnetic origin. As it will be demonstrated, the first order terms of this self-potential contain the relativistic energy (the rest and the kinetic energy) of the electrodynamic body, while higher order terms can be related to a new function, that can be correlated to the quantum potential. In this manner, we hope to provide a better understanding of quantum motion or, at least, to pave the way towards such an understanding. 

\section{The self-force}

\begin{figure}
\centering
\includegraphics[width=0.9\linewidth,height=0.5\linewidth]{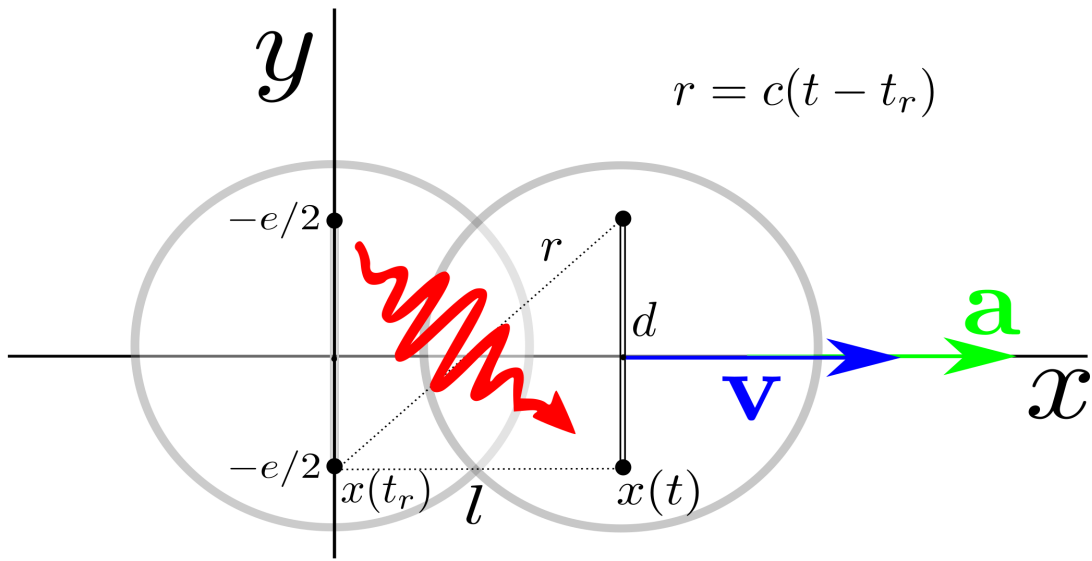}
\caption{\textbf{A model for an electrodynamic body}. An extended electron, modeled as a dumbbell joining two point charged particles (black dots) at a fixed distance $d$. The particle is shown at the retarded time $t_{r}$ and at a some later time $t$. During this time interval, the corpuscle accelerates in the x-axis, advancing some distance $l$ in such direction. As we can see, the particle in the upper part emits a field perturbation at the retarded time (red photon), and this perturbation reaches the second particle at the opposite part of the dumbbell at a later time (and vice versa). In this manner, an extended corpuscle can feel itself in the past. The speed and the acceleration of the particle are represented in blue and green, respectively.}
\label{fig:1}
\end{figure}

We begin with the Li\'enard-Wiechert potential \cite{lie98,wie01} for a body formed by two charged point particles attached to a neutral rod that move transversally along the x-axis. In general, any motion with transversal field component suffices to derive the main conclusions of this work. However, to avoid dealing with the rotation of the dumbbell, we restrict to a one-dimensional translational motion. This allows to keep mathematics as simple as possible, since the Li\'enard-Wiechert potential is retarded in time, and this non-conservative character of electrodynamics makes the computations very entangled. This elementary model was wisely designed in previous works to derive from first principles the Lorentz-Abraham force \cite{lo892,ab905} and also to study a possible electromagnetic origin of inertia \cite{gri83,gri89}. It is a toy model of an electron, represented as an extended electrodynamic body with approximate size $d$, as shown in Fig.~\ref{fig:1}. Among the aforementioned virtues, we also find that some properties resulting from considering more complex geometries (spherical, for example) of a particle, can be derived by superposition \cite{gri89}. We shall use this elementary model all along our exposition, which is more than sufficient to illustrate the fundamental mechanism that leads to electrodynamic fluctuations.

As we can see in Fig.~\ref{fig:1}, the first particle affects the other at a later time, since the perturbations of the field have to travel from one particle to the other. In other words, an extended body can affect itself. This sort of interaction is traditionally known as a self-interaction in the literature \cite{gri83} and, as can be seen ahead, for any charged particle, it produces an excitatory force, together with a recoil force and an elastic restoring force as well. The complete Li\'enard-Wiechert potential permits to write the electric field created by the first particle at the point of the second as
\begin{equation}
\bm{E}_{1}=\frac{q}{8\pi \epsilon_{0}}\frac{r}{(\bm{r}\cdot\bm{u})^3}\left(\bm{u}(1-\beta^{2})+\frac{1}{c^2}\bm{r}\times (\bm{u}\times\bm{a})\right),
\label{eq:1}
\end{equation}
where we have now defined the vector $\bm{u}=\bm{\hat{r}}-\bm{\beta}$, with the relative position between particles $\bm{r}(t_{r})$, their velocity $\bm{\beta}(t_{r})=\bm{v}(t_{r})/c$ and their acceleration $\bm{a}(t_{r})$ depending on the retarded time $t_{r}=t-r/c$. The retarded time appears due to the limited speed at which electromagnetic field perturbations travel in spacetime, according to Maxwell's equations \cite{max65}. This restriction imposes the constraint
\begin{equation}
r=c(t-t_{r}),
\label{eq:2}
\end{equation}
which assigns a particular time in the past from which the signals coming from one particle of the dumbbell affect the remaining particle.
As we shall see, the fact that dynamical systems under electrodynamic interactions are time-delayed (\emph{i.e.} the non-Markovian character of electrodynamics), is at the basis of the whole mechanism. Now we follow the picture in Fig.~\ref{fig:1} and write the position, the velocity and the acceleration vectors as $\bm{r}=l\bm{\hat{x}}+d\bm{\hat{y}}$,$\bm{\beta}=v/c\bm{\hat{x}}$ and $\bm{a}=a \bm{\hat{x}}$, respectively, where the distance $l=x(t)-x(t_{r})$ between the present position of the particle and the position at the retarded time has been introduced. Using these relations, the vector $\bm{u}$ can be computed immediately as
\begin{equation}
\bm{u}=\frac{(l-r\beta)\bm{\hat{x}}+d\bm{\hat{y}}}{r},
\label{eq:3}
\end{equation}
which, in turn, allows to write the inner product $\bm{r}\cdot\bm{u}=r-l\beta$, by virtue of the Pythagoras' theorem $r^{2}=(x(t)-x(t_{r}))^2+d^2$. Concerning the radiative fields, we can express the triple cross-product as $\bm{r}\times(r\bm{u}\times\bm{a})=-d^2 a \bm{\hat{x}}+d a l \bm{\hat{y}}$. We now compute the net self-force on the particle's centre of mass as
\begin{equation}
\bm{F}_{\rm{self}}=\frac{q}{2} (\bm{E}_{1}+\bm{E}_{2})=q E_{1x} \bm{\hat{x}},
\label{eq:4}
\end{equation}
where $\bm{E}_{2}$ is the force of the second particle on the first. Note that we have assumed that all the forces on the y-axis cancel, because we have simplified the model by using a rigid dumbbell to keep the distance of the charges fixed. This includes repulsive electric forces and also magnetic attractive forces as well. Therefore, in the present section we do not cover the much more complicated problem of the particle's stability, which is discussed in the last section of the present work. Such a problem is of the greatest importance and lead to the introduction of Poincar\'e's stresses in the past \cite{po905} and, among other reasons (\emph{e.g.} atomic collapse), to the rejection of classical electrodynamics as a fundamental theory \cite{abr04}. If prefered, from a theoretical point of view, the reader can consider that the two point particles of our model are kept at a fixed distance by means of some balancing external electromagnetic field oriented along the y-axis.

Now, we replace the value of the charge with the charge of the electron $q = -e$ to finally arrive at the mathematical expression describing the self-force of the particle, which is written as
\begin{equation}
\bm{F}_{\rm{self}}=\frac{e^2}{8\pi \epsilon_{0}} \frac{1}{(r-l \beta)^3} \left( (l-r \beta)(1-\beta^{2})-\frac{d^2}{c^2}a \right) \bm{\hat{x}}.
\label{eq:5}
\end{equation}

\section{The equation of motion}

We are now committed to write down Newton's second law in the non-relativistic limit $\bm{F}_{\rm{self}} = m \bm{a}$ and redefine the mass of the particle since, as we show right ahead, the electrostatic internal interactions add a term to the inertial content of the particle. The main purpose of the following lines is to expand in series the self-force to show its different contributions to the equation of motion. The two most resounding terms are the Lorentz-Abraham force and the force of inertia. However, we draw attention to other relevant nonlinear terms, which are of fundamental importance. These expansions will enable a discussion about the electromagnetic origin of mass and, based on such line of reasoning, we shall derive the appropriate and precise equation of motion.

As it has been shown in previous works \cite{gri83,gri89}, it is possible to express $l$ as a function of $r$ by means of the series expansion
\begin{equation}
l=x \left(t_{r}+\frac{r}{c} \right)-x \left(t_{r} \right)=\beta r+\frac{a}{2 c^2}r^2+\frac{\dot{a}}{6 c^3}r^3+\frac{\ddot{a}}{24 c^4}r^4+...
\label{eq:6}
\end{equation}
This trick of approximating magnitudes presenting delay differences by means of a Taylor series has been used sometimes in the study of delayed systems along history \cite{ai830}. We recall that this simplification is not a minor issue, since by truncating this expansion we are replacing a system with memory by a Markovian one. Nevertheless, the reader must be aware that delayed systems are infinite-dimensional. In fact, as we show below, any truncation of the previous equation is mistaken since, even though the time-delay $r/c$ is small, the terms in the acceleration, the jerk and so on, are not of order zero in such factor.

As shown in the Appendix, together with Eq.~(\ref{eq:2}), the previous expansion allows to express the corpuscle's size in terms of the time-delay by means of the series
\begin{equation}
d=r-\frac{a}{2 c^2}\beta r^2-\left(\frac{a^2}{8 c^4}+\beta \frac{\dot{a}}{6 c^3} \right)r^3+...
\label{eq:7}
\end{equation}
This Taylor series can be inverted to compute the expansion of $r$ in terms of $d$, which can be written to first order in $\beta$ as
\begin{equation}
r=d+\frac{a}{2 c^2}\beta d^2+\left(\frac{a^2}{8 c^4}+\beta \frac{\dot{a}}{6 c^3} \right)d^3+...
\label{eq:8}
\end{equation}
Finally, by inserting Eq.~(\ref{eq:8}) in the previous Eq.~(\ref{eq:6}) and then both equations in Eq.~(\ref{eq:5}), with the aid of Newton's second law, we compute, to first order in $\beta$, the identity
\begin{equation}
\left(m+\frac{e^2}{16 \pi \epsilon_{0}}\frac{1}{c^2 d} \right)\bm{a}=\frac{e^2}{8 \pi \epsilon_{0}} \left(\frac{1}{2 c^5}a^2 \bm{v}+\frac{5d}{16 c^6}a^2\bm{a}+\frac{1}{6 c^3}\dot{\bm{a}}+\frac{d}{24 c^4}\ddot{\bm{a}}+...\right),
\label{eq:9}
\end{equation}
after a great deal of algebra. These computations are enormously simplified by means of modern software for symbolic computation \cite{mat19}. 

We notice that the Lorentz-Abraham force has appeared in the third term of the right-hand side of this last equation, together with a few other linear and nonlinear terms. Interestingly, we recall that the term of inertia dominates all other terms for small speeds and accelerations. We can truncate this equation up to the jerk term $\dot{a}$, disregarding its nonlinearity and also derivatives of higher order. We can also define the renormalized mass of the electron as
\begin{equation}
m_{e}=m+\frac{e^2}{16 \pi \epsilon_{0}}\frac{1}{c^2 d},
\label{eq:10}
\end{equation}
and recall the relation between the electron's charge and Planck's constant by means of the fine structure constant
\begin{equation}
\hbar \alpha c=\frac{e^2}{4 \pi \epsilon_{0}},
\label{eq:11}
\end{equation}
according to Sommerfeld's equation \cite{som19}. Then, we get the approximated solution
\begin{equation}
\ddot{\bm{\beta}}-\frac{12 m_{e} c^2}{\hbar \alpha}\dot{\bm{\beta}} \left(1-\frac{5 \hbar \alpha d}{32 m_{e} c^3}\dot{\bm{\beta}}^2 \right)+\frac{3a^2}{c^2} \bm{\beta}+...=0,
\label{eq:12}
\end{equation}
which reminds of the equation of a nonlinear oscillator.

Thus, we see that the term of inertia, which is the linear term in the acceleration and which dominates when the particle is perturbed from rest, acts as an antidamping. This term is due to radiation fields and is responsible for the amplification of fluctuations. This fact does not contradict Newton's third law, since it is the addition of matter and radiation momentum that must be conserved as a whole. In other words, the particle can propel itself for a finite time by taking energy from its ``own" field. However, the nonlinear cubic term in $\dot{\beta}$ in Eq.~(\ref{eq:12}), which has opposite sign, limits the growth of the fluctuations. When the acceleration surpasses a certain critical value, the radiation reaction and the radiative fields do not act in phase anymore, and the fluctuations are damped away. Therefore, the pathological attributes that have been predicated of this marvelous recoil force \cite{gri89} are unjustified, and arise as a consequence of disregarding nonlinearities, which are responsible for the system's stabilization and, as we shall demonstrate, its self-oscillatory dynamics.

Importantly, at this point we notice that, if we assume that the inertia of the electron has an exclusive electromagnetic origin and recall that the dumbbell is neutral ($m = 0$), all the mass must come from the charged points. Then, using the Eqs.~\eqref{eq:10} and \eqref{eq:11} we can write the mass as
\begin{equation}
m_{e}=\frac{\hbar \alpha}{4 d c},
\label{eq:13}
\end{equation}
which was obtained in previous works \cite{gri83} and gives an approximate radius of the particle $r_{e} = d / 2 = 3.52 \times 10^{-16} \rm{m}$. Except for a factor of eight due to the dumbbell's geometry, this value corresponds to the classical radius of the electron. In this manner, we do not need to introduce spurious elements (artificial mechanical inertia) in the theory of electromagnetism, and simply use the D'Alembert's principle instead of Newton's second law \cite{da743}. If desired, and to extol Newton's intuition, the second law of classical mechanics would be a conclusion of electromagnetism, which is the most fundamental of classical theories. What it is amazing is that Newton was capable of figuring it out without any knowledge on electrodynamics. However, this wonderment partly fades out if we bear in mind the unavoidable corollary. For if mass is of electromagnetic origin, the gravitational field must be a residual electromagnetic field. If we are willing to accept these two inextricable facts, inertia would just be an internal resistance or self-induction force produced by the field perturbations to the motion of the charged body, when an external field is applied. We tackle more deeply this issue in the colophon of this work. 

In summary, we believe that it is more appropriate to simply consider Newton's second law as a static problem $\bm{F}_{\rm{ext}}+ \bm{F}_{\rm{self}} = 0$. In our case, we simply have $\bm{F}_{\rm{self}} = 0$. This way of posing the problem can be regarded as computing the geodesic equation of motion of the particle, as it occurs, for example, in the theory of general relativity. The resulting equation of motion reads
\begin{equation}
\left(1-\frac{v^2(t_{r})}{c^2} \right) \left(x(t)-x(t_{r})-\frac{r}{c} v(t_{r}) \right)-\frac{d^2}{c^2}a(t_{r})=0,
\label{eq:14}
\end{equation}
where we recall that for $v=c$ the first term vanishes, not allowing the particle to overcome the speed of light.

We now derive two relations that shall prove of great assistance in forthcoming sections to compute exact results. For this purpose, we use again the Pythagoras' theorem $r^{2}=(x(t)-x(t_{r}))^2+d^2$ and the equality appearing in Eq.~(\ref{eq:14}). By combining these two equations it is straightforward to derive a second order polynomial in $r$, which is solved yielding
\begin{equation}
r=\gamma d \sqrt{1+\gamma^6\dot{\beta}^2 \left(\frac{d}{c}\right)^2}+\gamma^{4} c \beta \dot{\beta}\left(\frac{d}{c}\right)^2,
\label{eq:15}
\end{equation}
where the Lorentz factor $\gamma = (1 - \beta^{2})^{-1/2}$ has been introduced and the kinematic variables are evaluated at the retarded time. Note that, contrary to the previous Eq.~\eqref{eq:8}, this expression is exact and has the virtue of suggesting that any consistent power series expansion of $r$ should be carried out in terms of the factor $d/c$. We also notice that, by virtue of this equation, the delay becomes dependent on the speed and the acceleration of the particle. As the corpuscle speeds up, the self-signals come from earlier times in the past. In other words, the light cone of the corpuscle is dynamically evolving, and this evolution selects different signals coming from the past.

Finally, the insertion of this relation into the equation $r^2=l^2+d^2$ leads to the obtainment of $l$ as a function of $\beta$ and $\dot{\beta}$ in a closed form. Again, this avoids the use of an infinite number of derivatives. The final result can be written as
\begin{equation}
l=\sqrt{\gamma^2c^2\beta^2\left(\frac{d}{c}\right)^2+\gamma^{8}c^{2}\dot{\beta}^{2}(1+\beta^{2})\left(\frac{d}{c}\right)^4+2 c^2 \gamma^{5} \beta \dot{\beta}\left(\frac{d}{c}\right)^3 \sqrt{1+\gamma^6\dot{\beta}^2\left(\frac{d}{c}\right)^2}}.
\label{eq:16}
\end{equation}

These two Eqs.~(\ref{eq:15}) and (\ref{eq:16}) will allow us to derive analytical results in a fully relativistic manner, specially concerning the self-potential.

\section{The instability of rest}

Even though we shall prove a more general statement in Sec.~5, we believe that the fact that oscillatory dynamics can be the default state of matter, instead of a stationary state, is of paramount importance. In turn, this study provides a double check of the results presented in such section. Therefore, we independently study the stability of the rest state of the particle in the following lines. Our goal is to show that the rest state is unstable and to identify the magnitude that leads to the amplification of fluctuations. For this purpose, we begin with the expansion appearing in Eqs.~(\ref{eq:6}) and (\ref{eq:8}), and replace them in Eq.~(\ref{eq:14}), neglecting all the nonlinear terms. Such terms can be disregarded since the rest state is represented by $v$ and all its higher derivatives are equal to zero. Thus, when slightly perturbing the rest state of the charged particle, we only need to retain linear contributions. The resulting infinite-dimensional differential equation is
\begin{equation}
-\frac{1}{2 c^2 d}\bm{a}+\frac{1}{6 c^3}\dot{\bm{a}}+\frac{d}{24 c^4}\ddot{\bm{a}}+\frac{d^2}{120 c^5}\dddot{\bm{a}}+...=0.
\label{eq:17}
\end{equation}

This equation can be more clearly written as a Laurent series in the factor $d/c$, as previously suggested. We obtain the result
\begin{equation}
-\frac{1}{2}\frac{c}{d}\bm{a}+\frac{1}{6}\dot{\bm{a}}+\frac{1}{24}\frac{d}{c}\ddot{\bm{a}}+\frac{1}{120}\frac{d^2}{c^2}\dddot{\bm{a}}+...=0,
\label{eq:18}
\end{equation}
which can be generally expressed as
\begin{equation}
-\frac{1}{2}\bm{a}+\sum^{\infty}_{n=1}\frac{1}{(n+2)!}\frac{\textrm{d}^{n}\bm{a}}{\textrm{d}t^n} \left( \frac{d}{c} \right)^n=0.
\label{eq:19}
\end{equation}

The characteristic polynomial of this equation is obtained by considering as solution $a(t) = a_{0} e^{\lambda t}$. We compute the relation
\begin{equation}
-\frac{1}{2}+\sum^{\infty}_{n=1}\frac{1}{(n+2)!} \left( \frac{\lambda d}{c} \right)^{n}=0,
\label{eq:20}
\end{equation}
which can be more elegantly written by using the Maclaurin series of the exponential function. If we redefine it by means of the variable $\mu = \lambda d/c$, we get
\begin{equation}
-\frac{1}{2}+\frac{1}{\mu^2}\sum^{\infty}_{n=1}\frac{\mu^{n+2}}{(n+2)!}=-\frac{1}{2}+\frac{1}{\mu^2}\left( e^{\mu}-\frac{\mu^2}{2}-\mu-1\right)=0.
\label{eq:21}
\end{equation}

\begin{figure}
\centering
\includegraphics[width=1.0\linewidth,height=0.5\linewidth]{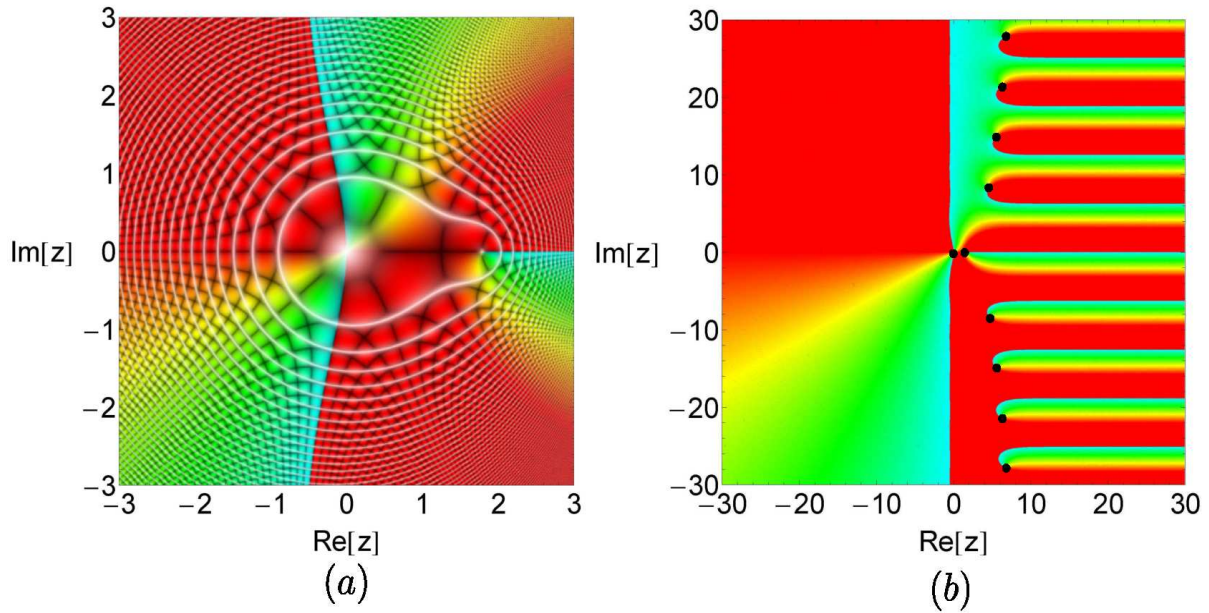}
\caption{\textbf{The roots of the polynomial $f(z)=z^2+z+1-e^z$}. (a) A domain coloring representation of the function. The color represents the phase of the complex function. The shiny level curves represent the values for which $|f(z)|$ is an integer, while the dark stripes are the curves Re$f(z)$ and Im$f(z)$ equal to a constant integer. The roots and poles can be localized where all colors meet. In the present case we clearly identify the roots $z=0$ and $z=9/5$. (b) Here a zoom out of the function is shown, with the distribution of zeros (black dots). The coloring scheme has been simplified. As can be seen, all of them are distributed on the positive real part of the complex plane.}
\label{fig:2}
\end{figure}

The solutions to this equation can be obtained numerically. Apart from zero, the only purely real solution can be nicely approximated as
\begin{equation}
\lambda=\frac{9}{5}\frac{c}{d},
\label{eq:22}
\end{equation}
which is a positive value. In summary, the rest state is not stable in the Lyapunov sense \cite{lya92}, and this implies that the particle can not be found at rest. In fact, as can be shown in Fig.~\ref{fig:2}, the complex function $f(z)=z^2+z+1-e^{z}$ has an infinite set of zeros in the complex plane. All of them have a positive real part, while all except two of them are complex conjugate numbers with non-zero imaginary part. It can be analytically shown that, for zeros with negative real part to exist, they have to be confined in a small region close to the origin. Consequently, numerical simulation is enough to confirm both the instability of rest and the existence of self-oscillations in the system. 

As more generally stated below, everything is jiggling because electromagnetic fluctuations are amplified. Consequently, motion would be the essence of being and not rest, as could be inferred from the principle of inertia in Newtonian mechanics. More precisely, and as we are about to show, it is uniform motion that it is unstable. This notion is precisely a strong suggestion in order to assume that inertia has an electromagnetic origin. But we shall give a more compelling one below. Be that as it may, the instability of stillness can be considered, by far, the most fundamental finding of the present analysis.

\section{Self-oscillations}

We now proceed to show the existence of limit cycle oscillations of the particle. Since the rest state is unstable and the speed of light can not be surpassed according to Eq.~(\ref{eq:14}), the only possibilities left are uniform motion or some sort of oscillatory dynamics, weather regular or chaotic. In the first place, we rewrite the Eq.~(\ref{eq:14}) to a more amenable and familiar form. We have
\begin{equation}
\frac{d^2}{c^2}a(t_{r})+\frac{r}{c} \left(1-\frac{v^2(t_{r})}{c^2} \right) v(t_{r})+\left(1-\frac{v^2(t_{r})}{c^2} \right) \left(x(t_r)-x(t)\right)=0.
\label{eq:23}
\end{equation}

The main handicap of this equation is that it is expressed in terms of the retarded time $t_r = t - r / c$, which it is the customary expression of the Li\'enard-Wiechert potentials. To obtain the same equation in terms of the present time $t$, we simply perform a time translation to the advanced time $t_a = t + r /c$. This allows to write
\begin{equation}
a(t)+\frac{r}{d} \frac{c}{d} \left(1- \frac{v^2(t)}{c^2} \right) v(t)+\left(\frac{c}{d}\right)^2 \left(1-\frac{v^2(t)}{c^2} \right) \left(x(t)-x \left( t+\frac{r}{c} \right) \right)=0.
\label{eq:24}
\end{equation}

But now the problem is that this equation depends on the advanced time. In other words, Eq.~(\ref{eq:24}) allows to derive the position and velocity at some time from the knowledge of such position and velocity in the past, by using the position in the future. This equation reminds of the equation of a self-oscillator \cite{jen13}. Apart from the term of inertia and the linear oscillating term representing Hooke's law \cite{hoo78}, we have two nonlinear contributions. On the one hand, the second contribution on the left hand side acts here as a damping term and it is responsible for the system's dissipation. This term is identical to other terms appearing in traditional self-oscillating systems, as for example the oscillator introduced by Lord Rayleigh's to describe the motion of a clarinet reed \cite{ray45} and, to some extent, also to the Van der Pol's oscillator \cite{van20}. On the other hand, the antidamping comes from the advanced potential. At first sight, in the limit of small velocities, the frequency of oscillation is $\omega_{0} = c/d$, what allows to approximate the period as
\begin{equation}
T=4 \pi \frac{r_{e}}{c},
\label{eq:26}
\end{equation}
where $r_{e}=d/2$ is the radius of the electron. This equation gives a value of the period of approximately $T = 1.18 \times 10^{-22} \rm{s}$ for the classical radius of the electron. Therefore, the particle would oscillate very violently, giving rise to an apparently stochastic kind of motion. This motion and the value of the frequency should not be unfamiliar to quantum mechanical theorists, since they can be related to the trembling motion appearing in Dirac's equation \cite{sch30}, commonly known as \emph{zitterbewegung}. 

As we have shown in Sec.~2, the time-delay $r$ depends on the kinematic variables. We insist that, in this sense, despite of the simplicity of the model at analysis, we are facing a terribly complicated dynamical system, since the delay itself depends on the speed and the acceleration of the particle. This kind of systems are formally referred in the literature as state-dependent delayed dynamical systems \cite{sie17} and, from an analytical point of view, they are mostly intractable. Importantly, we note that for a system of particles, the dependence of the delay of a certain particle on the kinematic variables of the others at several times in the past and at the present as well, turn electrodynamics into a nonlocal theory \cite{jac02}. This functional dependence sheds some light into the significance of entanglement, which can now be regarded as a process of entrainment of nonlinear oscillators \cite{pan08}.

All this complexity notwithstanding, since we just aim at illustrating the existence of self-oscillatory dynamics, we shall have no problems concerning the integration of this system. According to Eq.~(\ref{eq:22}), when the system is amplifying fluctuations from its rest state, we see that the rate at which the amplitude of fluctuations grows is comparable to the period of the oscillations. Therefore, averaging techniques, as for example the Krylov-Bogoliubov method \cite{kry35}, cannot be safely applied in the present situation. More simply, we consider the differential equation (\ref{eq:24}) and write it in the phase space as
\begin{align}
&\dot{x} =y, \nonumber \\
&\dot{y} =-\frac{c}{d}\frac{r}{d}\left(1-\frac{y^2}{c^2} \right)y-\left(\frac{c}{d}\right)^2\left(1-\frac{y^2}{c^2} \right)\left(x-x_{\tau} \right), \label{eq:27} 
\end{align}
where $x_{\tau}$ represents the position at the advanced time $t+\tau=t+r/c$. As we have shown in the previous section, the fixed point $\dot{x} = \dot{y} = 0$ is unstable. Apart from the rest state, asymptotically, there can be only two possibilities. Since the speed of light is unattainable for massive particles, either the particle settles at a constant uniform motion with a lower speed, or its speed fluctuates around some specific value. We do not enter into the issue weather these asymptotic oscillations are periodic, quasiperiodic or chaotic. We shall just prove that uniform motion is not stable and, consequently, self-oscillatory dynamics is the only possibility, whatever its periodicity might be. Assume that uniform motion is possible at some speed $y$, which is a constant number $\beta c$. Then, we have that $x(t) = y t$ and also that $x(t + r / c) = yt + yr /c$, which implies $x - x_{\tau} = -yr /c$. Substitution in Eq.~(\ref{eq:26}) yields
\begin{align}
&\dot{x} =y,\nonumber \\
&\dot{y} =-\frac{c}{d}\frac{r}{d}\left(1-\frac{y^2}{c^2} \right)y+\frac{c}{d}\frac{r}{d}\left(1-\frac{y^2}{c^2} \right)y=0. \label{eq:28} 
\end{align}

Thus, certainly, any uniform motion is also an invariant solution (a fixed trajectory, so to speak) of our state-dependent delayed dynamical system. However, it is immediate to show that this solution is unstable as well. We prove this assertion by computing the variational equation related to inertial observers
\begin{align}
&\delta\dot{x} =\delta y,\nonumber \\
&\delta \dot{y} = -\frac{c}{d}\frac{\delta r}{d}\left(1-\frac{y^2}{c^2}\right)y-\frac{c}{d}\frac{r}{d}\left(1-\frac{y^2}{c^2}\right)\delta y+\frac{c}{d}\frac{r}{d}\frac{2 y^2}{c^2}\delta y-\nonumber \\ 
&~~~~~~-\frac{c}{d}\frac{r}{d}\frac{2 y^2}{c^2} \delta y-\left(\frac{c}{d}\right)^2\left(1-\frac{y^2}{c^2} \right)\left(\delta x-\delta x_{\tau} \right). 
\label{eq:29} 
\end{align}
At this point, we have to compute $\delta r$ at $\dot{y} = 0$ and $y = \beta c$, with $\beta$ a constant value. Using the formula (\ref{eq:15}), but evaluated at the present time, this calculation can be carried out without difficulties yielding
\begin{equation}
\delta r(t)=\gamma^4 \beta \left(\frac{d}{c}\right)^2 \delta \dot{y}(t)+d \delta \gamma(t),
\label{eq:30}
\end{equation}
where again we notice that the variables are evaluated at the present time. Gathering terms and using the fact that $r = \gamma d$ for $\dot{y} = 0$, we finally arrive at the variational problem
\begin{align}
&\delta\dot{x} =\delta y,\nonumber \\
&\delta \dot{y} \gamma^2 = -\frac{c}{d}\gamma \delta y-\left(\frac{c}{d}\right)^2 \left(1-\beta^2 \right) \left(\delta x-\delta x_{\tau} \right).\label{eq:31} 
\end{align}

\begin{figure}
\centering
\includegraphics[width=0.6\linewidth,height=0.5\linewidth]{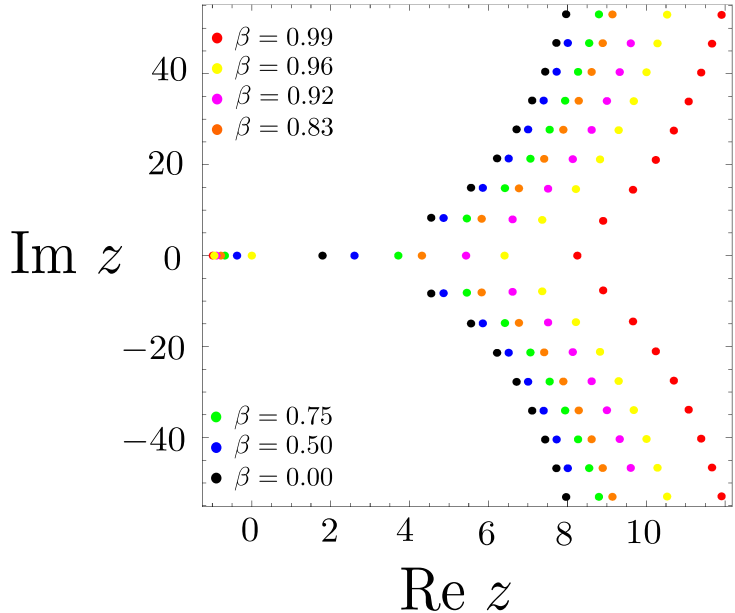}
\caption{\textbf{The roots of the polynomial $f(z)=z^2+z+(1-e^z)(1-\beta^2)$}. The complex roots of the $f(z)$ have been numerically computed using Newton's method for different values of the speed, ranging from the rest state $(\beta=0)$ to the ultrarelativistic limit. As we can see, the values of the imaginary part do not seem to depend on $\beta$ and can be written as multiples of a fundamental frequency. Since $z=\gamma d/c$, we get the spectrum of frequencies for the self-oscillation $\omega_{n} \propto n c/\gamma d$, at least right after the state of uniform motion is slightly perturbed.}
\label{fig:3}
\end{figure}

If we consider solutions of the form $ \delta x = A e^{\lambda t}$, the characteristic polynomial of the system (\ref{eq:31}) is found. It reads
\begin{equation}
\lambda^2 \gamma^2+\frac{c}{d} \gamma  \lambda+\left(\frac{c}{d}\right)^2(1-\beta^2)(1-e^{\lambda \gamma d/c})=0.
\label{eq:32}
\end{equation}
Two limiting situations appear. In the non-relativistic limit $\beta \rightarrow 0$ we can write
\begin{equation}
\lambda^2 +\frac{c}{d}  \lambda+\left(\frac{c}{d}\right)^2(1-e^{\lambda d/c})=0.
\label{eq:33}
\end{equation}
which, considering $ \mu = \lambda d/c$, can be written as
\begin{equation}
\mu^2 +\mu+1-e^{\mu}=0.
\label{eq:34}
\end{equation}
This is in conformity with previous results (see Eq.~(\ref{eq:21})). Finally, in the relativistic limit, we get
\begin{equation}
\mu^2 +\mu+(1-e^{\mu})(1-\beta^2)=0,
\label{eq:35}
\end{equation}
where we have now defined $\mu =\lambda \gamma d/c$. Except for one eigenvalue, the real part of the solutions to this equation are always positive and therefore unstable for any value of $\beta$, as confirmed by numerical simulations (see Fig.~\ref{fig:3}). Again, an infinite set of frequencies are obtained, which can be written as
\begin{equation}
\omega_{n}=\eta_{n} \frac{c}{\gamma d},
\label{eq:36}
\end{equation}
where the factor $\gamma$ accounts for the time dilation related to Lorentz boosts. The parameters $\eta_{n}$, according to Fig.~\ref{fig:3}, can be reasonably approximated by means of a linear dependence on $n$, which is an integer greater or equal than one. From the same image we can see that these parameters are independent of the speed of the system.

In this manner, we have proved the existence of self-oscillating motion in this dynamical system for all values of $\beta$. We recall \emph{en passant} that the damping term and the delay introduce an arrow of time in the system \cite{mac03}. In other words, the limit cycle can be run in one time direction, but not in the reverse. This lack of reversibility is inherent to delayed systems, which depend on their previous history functions \cite{daz17} and, therefore, are fundamentally non-conservative systems. Nevertheless, we note that the violation of energy conservation should only last a small time until the invariant limit set is obtained, and that it applies as long as long as we just look at the particle and not to the fields. This fact evokes nicely the time-energy uncertainty relations, as can be noticed in the next section. Even though self-oscillations were pointed out a long time ago for a charged particle \cite{boh48}, the instability of ``classical" geodesic motion had been unnoticed before, perhaps due to the fact that artificial inertia was assumed and because there exists a dependence of the degree of instability on the geometry of the particle \cite{lop20}. This would be simply natural, given the complexity of retarded fields, and justifies the use of the apparently simple present model.

\section{The self-potential}

In the present section we obtain the relativistic expression of the potential energy of the charged body, starting again from the Li\'enard-Wiechert potential of the electromagnetic field. We denote this self-energy as $U$ since, it can be regarded as the non-dissipative energy required to assemble the system and set it at a certain dynamical state. As it will be clear at the end of the section, it harbors both the rest and the kinetic energy of the particle and also a kinematic formulation of what we suggest might be the quantum potential, which is frequently written as $Q$ in the literature \cite{noh06}.

The electrodynamic energy of the dumbbell can be computed as the energy
required to settle it in a particular dynamical state. Since the magnetic fields do not perform work, we would have to compute the integral
\begin{equation}
U=\frac{e}{2} \int^{\bm{r}}_{\bm{r}_{0}} \bm{E}\cdot \textrm{d}\bm{r}=-\frac{e}{2} \int^{\bm{r}}_{\bm{r}_{0}} \nabla \varphi \cdot \textrm{d} \bm{r}-\frac{e}{2} \int^{\bm{r}}_{\bm{r}_{0}} \frac{\partial \bm A}{\partial t}\cdot \textrm{d}\bm{r},
\label{eq:37}
\end{equation}
along some specific history describing a possible journey of the dumbbell. However, it can be shown that the second term is just the dissipative contribution. Therefore, we concentrate on the irrotational part of the field. The electrodynamic potential energy of the dumbbell is just given by the Li\'enard-Wiechert potential as
\begin{equation}
U=\frac{e^2}{16 \pi \epsilon_{0}}\frac{1}{\bm{r} \cdot \bm{u}},
\label{eq:38}
\end{equation}
where the additional one fourth factor comes from the fact that each charge brings a value $q =- e/2$. This can be written by means of the Eq.~($\ref{eq:3}$) as
\begin{equation}
U=\frac{\hbar \alpha c}{4(r-l \beta)}.
\label{eq:39}
\end{equation}
If we now substitute the Eqs.~(\ref{eq:15}) and (\ref{eq:16}), and develop them in powers of $d/c$, we obtain the series expansion of the self-potential
\begin{equation}
U=\gamma \frac{\hbar \alpha c}{4d}-\gamma^7 \frac{a^2}{2 c^2}\frac{\hbar \alpha}{4}\left(\frac{d}{c}\right)+\gamma^{13} \frac{3 a^4}{8 c^4}\frac{\hbar \alpha}{4}\left(\frac{d}{c}\right)^3-\gamma^{19} \frac{5 a^6}{16 c^6}\frac{\hbar \alpha}{4}\left(\frac{d}{c}\right)^5+...
\label{eq:40}
\end{equation}
We recall that these computations are very lengthy and again strongly recommend the use of software for symbolic computation. We arrive in this manner at the crucial point of this exposition. If we once again simply assume the idea that inertia has an electromagnetic origin, we can write the size of the particle as
\begin{equation}
d=\frac{\hbar \alpha}{4 m_{e} c}.
\label{eq:41}
\end{equation}
Substitution in the previous equation yields the series
\begin{equation}
U=\gamma m_{e}c^2-\frac{\hbar^2}{2 m_{e}} \frac{\alpha^2}{8 c^2} \gamma \left( \gamma^6\frac{a^2}{2 c^2}-\gamma^{12} \frac{3 a^4}{8 c^4}\left(\frac{d}{c}\right)^2+\gamma^{18} \frac{5 a^6}{16 c^6}\left(\frac{d}{c}\right)^4-... \right),
\label{eq:42}
\end{equation}
which can be written more formally as
\begin{equation}
U=\gamma m_{e}c^2+\frac{\hbar^2}{2 m_{e}} \frac{\alpha^2}{32 r_{e}^2} \gamma \sum_{n=1}^{\infty}{q_{n}(-1)^n\gamma^{6n} \frac{a^{2n}}{c^{2n}}}\left(\frac{d}{c}\right)^{2n},
\label{eq:43}
\end{equation}
where the coefficients $q_{n}=\{1/2,3/8,5/16,35/128,63/256...\}$ of the expansion belong to a sequence, which can be computed from the quadrature
\begin{equation}
q_{n}=\int^{1}_{0} \cos^{2n}(2\pi x) \textrm{d} x=\frac{(2n-1)!!}{2^n n!}.
\label{eq:44}
\end{equation}

We clearly identify two terms in Eq.~(\ref{eq:43}). The first one is just the relativistic energy \cite{ein05}, which contains both the rest and the kinetic energy of the particle. But note that, in addition to the kinetic and the rest energy of the particle, the potential
\begin{equation}
Q=\frac{\hbar^2}{2 m_{e}} \frac{\alpha^2}{32 r_{e}^2} \gamma \sum_{n=1}^{\infty}{q_{n}(-1)^n\gamma^{6n} \frac{a^{2n}}{c^{2n}}}\left(\frac{d}{c}\right)^{2n},
\label{eq:45}
\end{equation}
has unveiled as a new contribution. By inserting the integral appearing in Eq.~(\ref{eq:44}) into Eq.~(\ref{eq:45}), we can derive, after summation of the series and one additional integration, the potential
\begin{equation}
Q=-\frac{\hbar^2}{2 m_{e}} \frac{\alpha^2}{32 r_{e}^2} \gamma \left(1-\frac{1}{\sqrt{1+\gamma^6 \dot{\beta}^2 \left( \frac{d}{c}\right)^2}} \right),
\label{eq:46}
\end{equation}
which vanishes for uniform motion. Again, we note how the Lorentz factor precludes traveling at speeds higher or equal than the speed of light.

This potential evokes nicely the quantum potential appearing in Bohmian mechanics \cite{boh52,boh522}, with the same term $\hbar^2/2m_{e}$ preceding it. Importantly, we notice the dependence of fluctuations on the fine structure constant. Moreover, we have found a dependence of this potential on the acceleration of the particle that, we should not forget, is evaluated at the retarded time. On the other hand, since
\begin{equation}
Q=-\frac{\hbar^2}{2 m_{e}}\frac{\nabla^2 R}{R},
\label{eq:47}
\end{equation}
in quantum mechanics, we can settle a bridge between the square modulus of the wave function and the kinematics of the particle in the non-relativistic limit. In this way, we would restore the old relationship between forces and geometrical magnitudes. Once the dynamics is constrained to the asymptotic limit cycle, a relation between the acceleration of the particle and its position can be established and replaced in $Q$. Then, the resulting partial differential equation is similar to Helmholtz's equation
\begin{equation}
\nabla^2 R+\frac{2 m_{e}}{\hbar^2}Q{R}=0,
\label{eq:48}
\end{equation}
while we can independently write down the Hamilton-Jacobi equation for a particle
immersed in an external potential $V(x,t)$. In the non-relativistic limit, it is given by
\begin{equation}
\frac{\partial S}{\partial t}+\frac{1}{2 m_{e}}(\nabla S)^2+Q+V=0.
\label{eq:49}
\end{equation}

In principle, once the two previous Eqs.~(\ref{eq:48}) and (\ref{eq:49}) have been solved using the knowledge of the trajectory of the particle, the wave function can be built as
\begin{equation}
\psi(x,t)=R(x,t)\exp\left({\frac{i}{\hbar} S(x,t)}\right),
\label{eq:50}
\end{equation}
even though this solution may not be easily attained in most cases, specially when an external potential is present. Interestingly, we can see from these relations that the wave function is a real objective field, as claimed in the seminal works of David Bohm \cite{boh52,boh522}, and not just a probabilistic entity. Both its modulus and phase are related to internal and external electrodynamic forces.

To gain some insight into the self-potential of the ``free" particle, we illustrate these ideas by means of an example. For this purpose, we can invoke the oscillatory dynamics after the transient amplification to show the repulsive nature of electrodynamic fluctuations. A conservative version of the potential $Q_{c}(x)$ can be derived, which should only be regarded as an illustrative approximation. If we disregard the delay and consider the approximation $a = -\omega^2_{0} x$, in the non-relativistic limit, and keeping just the two first term of the series, we obtain the potential
\begin{equation}
Q_{c}(x)=-\frac{\hbar^2}{2 m_{e}}\frac{\alpha^2}{64 r_{e}^2} \left( \frac{1}{d^2}x^2-\frac{3}{4d^4}x^4\right).
\label{eq:51}
\end{equation}

This potential is very well known in the world of nonlinear dynamical systems, since it appears in the Duffing oscillator \cite{duf18}. This oscillator has been a paradigmatic model in the study of chaotic dynamical systems and has received remarkable attention both in physics and engineering, since it can describe many important phenomena, such as beam buckling, superconducting Josephson parametric amplifiers, or ionization waves in plasmas, among many others. It illustrates in a very clear manner the instability of stillness, because $Q_{c}(x)$ presents a maximum at $x=0$. In particular, this potential is responsible for the spontaneous symmetry breaking of the Poincar\'e group. We recall that symmetry breaking is a typical feature of nonlinear dynamical systems \cite{pri71,nic71}.

Interestingly, this potential can be written in a simplified form as
\begin{equation}
Q_{c}(x)=-\frac{1}{2}\hbar \omega \left( \frac{1}{2d^2}x^2-\frac{3}{8 d^4}x^4 \right),
\label{eq:52}
\end{equation}
where the frequency ${\omega}=\alpha c/2d$ has been defined, which is manifestly related to the frequency of \emph{zitterbewegung} of the dumbbell. 

What we find of the greatest interest about this expression is that it nicely evokes Planck's relation. Moreover, we recall that $m_{e}$ is proportional to $\hbar$, as long as we are in a position to assume that mass is of electromagnetic origin. Therefore, all sorts of energy and momentum can be ultimately written as proportional to Planck's constant. For example, the rest energy of the electron is written as $\hbar \omega/2$. It is then reasonable to argue that photons, which are light pulses emitted from accelerated electron transitions between different energy states, have energy $E=\hbar \omega$. Furthermore, by considering the relativistic relation $E = p c$, it is immediate to obtain from this equality that $p=\hbar k$, which brings in the De Broglie's relation between momentum and wavelength.
\begin{figure}
\centering
\includegraphics[width=1.0\linewidth,height=0.5\linewidth]{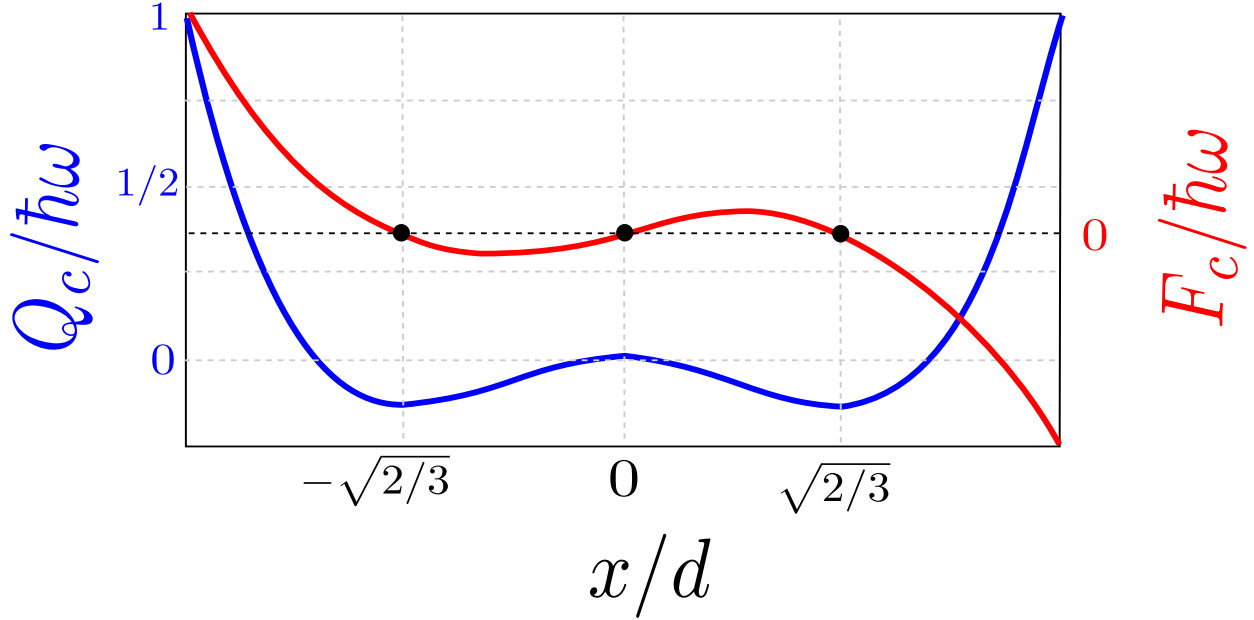}
\caption{\textbf{The quantum potential $Q_{c}(x)$}. This conservative approximation of the repulsive potential (blue line) has an unstable fixed point at the origin $x^{*}=0$, flanked by two minima, representing stable fixed points at $x^{*}=\pm\sqrt{2/3}$. The repulsive character of this potential guarantees the perpetual oscillatory motion of electrodynamic bodies. An approximation of the self-force is shown in red.}
\label{fig:4}
\end{figure}

As we can see, perhaps the main problem when studying the electrodynamics of extended bodies is that it leads to very complicated state-dependent delayed differential equations. Things would get terribly complicated if continuous bodies are considered, instead of the simple toy discrete model used here \cite{lop20}. This physical phenomenon arises as a consequence of the principle of causality, which imposes a limited speed at which information can travel in physics, introducing an infinite number of degrees of freedom in the nonlinear Lagrange equations. In fact, we wonder how the principle of least action can be reformulated to cover the complex time-delayed systems appearing in electrodynamics. In light of these facts, and from a practical point of view, the Schr\"odinger equation \cite{sch26} would be surely a much more appropriate and manageable mathematical framework than the use of the complicated functional differential equations resulting from the Li\'enard-Wiechert potentials to treat quantum problems. Certainly, it would not be surprising that partial differential equations, which have an infinite number of degrees of freedom, are of so much usefulness replacing delayed systems, which harbor an infinite number of degrees of freedom as well.

\section{Discussion}

As we have shown, the dynamics of an extended charged moving body has resemblances with the dynamics of the silicon droplets experimentally found in the recent years. However, in our picture, the waves travelling with the particle ``belong" to the particle itself, and do not require of any medium of propagation (any aether), since they are of electromagnetic origin. In our model, the fluctuations arise as self-interactions of the particle with its own field and have as analogy the fluctuating platform appearing in their experiments \cite{cou05}. Nevertheless, this analogy must be drawn with great care, since the physical phenomenon leading to fluctuations in our moving charged body is not resonance, but self-oscillation \cite{jen13}. In particular, we predict a simple relation of proportionality between quantum fluctuations and the coupling electromagnetic constant $\alpha$. Concerning self-oscillations, we also recall that a nonlocal probabilistic theory equivalent to a conservative diffusion process has been developed not so long ago, which is mathematically equivalent to non-relativistic quantum mechanics \cite{nel85}. This is in agreement with the present work since, as we have shown, our corpuscle exhibits very violent oscillations, as it is also suggested in other works \cite{boh52,boh522}.

The most astonishing consequence of the present work is the demonstration of the possibility of an instability of natural or uniform motion, which defies common intuition and beliefs on radiation as a purely damping field on electromagnetic extended moving sources. We believe that this misunderstanding is present in the beginning of many important introductory texts on quantum theory to justify the imperious necessity of a quantum mechanical theory that has no basis on the classical world \cite{dir81}. On the contrary, the present work suggests that self-interactions provide the required repulsive force (the quantum force) to avoid the collapse of electrodynamical systems. In particular, we predict that self-interactions and recoil forces are enough to stabilize the hydrogen atom and prevent its collapse \cite{raj04}. This is because radiation has, not only stabilizing dissipative effects as a whole on the system, but antidamping effects as well through self-excitation and radiation reaction on its several components. In the same way that it can excite an electron inside an atom to higher energy levels, it can self-excite an extended object by self-absorption. 

We also note that the wave-particle duality is immediately solved in our framework. The waves are just perturbations of the fields, and any charged accelerated particle can present such perturbations as a consequence of its self-oscillatory dynamics. Furthermore, there does not exist a fundamental particle that does not participate from some fundamental interaction and, consequently, there can be a pilot-wave \cite{deb27} attached to any charged particle in accelerated motion. Importantly, we highlight the rich dynamical feedback interaction between these two apparently differentiated entities. We recall that feedback is a crucial phenomenon for the understading of nonlinear dynamical systems in general, chaotic dynamics and, specially, for control theory \cite{wie48}. In light of this paragraph, it seems obvious that nothing can travel faster than field perturbations since, any aggregate of charge, whatever its nature is, will show resistance to acceleration due to its electromagnetic energy. This intuition brings back the concept of \emph{vis insita}, as appearing in Newton's work \cite{new99}. A concept that is also related to the original notion of inertia and Galileo's \emph{resistenza interna} \cite{gal32}, and which can be traced back to the seminal works of the dominic friar Domingo de Soto \cite{dom55,mir09}.

We now bring to discussion the most delicate point of the present work. The fact that the inertia of a body might be of electromagnetic origin (electroweak and strong, if desired) is and old argument in physical theories. As we have shown, it has been a sufficient and necessary condition to derive Newton's second law, kinetic energy, Einstein's mass-energy relation and what seems to be the quantum potential, just from Maxwell's electrodynamics. In this way, the present work gives a foundation of classical and quantum mechanics in the theory of electrodynamics  \cite{lyl10}. Perhaps, the greatest lesson of Einstein's relation is not that energy is mass, but that mass is a useful and simple way to gather the constants appearing in electrostatic energy. Consequently, we shall not invoke Occam's razor to defend the idea of gravitational mass as a redundant concept in fundamental physics. Instead, we adopt a more prudent position and focus the attention on the fact that our findings imply to reconsider Newton's second law as a law of statics, just as suggested by D'Alembert. In light of these facts, we believe that it is very natural that an electrodynamic mechanism gives mass to fundamental particles in the standard model, which is luckily known
nowadays thanks to the work of Higgs \cite{hig64}.

Following the same line of reasoning, this idea would perfectly connect with the theory of general relativity, since the principle of equivalence simply states that, in a non-inertial reference frame comoving with a body, any object experiences forces of inertia. In fact, these forces are equivalent to a gravitational field. Therefore, an electromagnetic theory of the gravitational field would also be in accordance with the principle of equivalence. Moreover, the identity of inertial and gravitational mass would be the consequence of a very simple fact, $i.e.$, their common electromagnetic origin. However, we must be careful at this point, since electromagnetic forces create strong ripples in space-time. Thus, a free falling extended charged particle in a gravitational field should experience very strong tidal self-forces. As we have shown, these forces can lead to self-oscillations. 

Delving deeper into the principle of covariance, we recall that the electromagnetic stress-energy tensor can be plugged into Einstein's equation and interpreted as a curvature of spacetime. The Einstein-Maxwell equations are terribly nonlinear high-dimensional partial differential equations, which can have as solutions solitary waves \cite{ale80,fab12,mis57}. Certainly, the model presented in this work is far too simplistic and unrealistic, because it assumes a rigid solid as a particle, which is contrary to electromagnetic theory, and whose structure is unstable. We expect particles to rotate and also to be deformable, and wonder if these two properties should be enough to stabilize the electron.

In this framework, gravitational waves would simply emerge from light waves. As a matter of fact, if the force of gravitation had an electromagnetic origin, the gravitational field, as a residual field, would have to be much weaker, which it is well-known to be the case. The fact that it falls with an inverse-square law should not be \emph{a priori} regarded as a problem. In fact, an average inverse quadratic law can be derived from radiative fields of a system of oscillating particles, which originally fall with the inverse of the distance. However, as far as the author has investigated, deriving a precise relation between the gravitational constant $G$ and the electron's charge $e$ from the Li\'enard-Wiechert potential of a system of particles would remain an open problem of paramount relevance.

To conclude, we would also like to evince our most radical skepticism concerning the present analysis. Firstly, the simplicity of the model should prevent us from drawing too general conclusions. 
It can be shown that purely longitudinal motion of the dumbbell is dissipative. Although this motion by itself is unstable to transverse perturbations, the authors recognize to have found a dependence of instability on the geometry of an electrodynamic moving body \cite{lop20}. As the shape of the body turns from oblate to prolate, a Hopf bifurcation befalls. Therefore, it might happen that some external electromagnetic field is necessary to unleash the oscillation for more complicated bodies. Or, perhaps, the rotational motion of the particle is essential to have unstable dynamics independently of its geometry. Secondly, a full correspondence between electrodynamics and the relativistic formalism of quantum mechanics has not been here provided. Nevertheless, and to close this lengthy discussion, we hope that this new perspective, based on modern theories of nonlinear dynamics, might serve to enlighten the complex dynamics of elementary classical particles and, if not, at least to drive physics closer to the establishment of a dynamical picture of fundamental particles, if such an endeavor is allowed and possible.

\section{Acknowledgments}
The author wishes to thank Alexandre R. Nieto for valuable comments on the elaboration of the present manuscript and discussion on some of its ideas. He also wishes to thank Alejandro Jenkins and Juan Sabuco for introducing him to the key role of self-oscillation in open physical systems, and for fruitful discussions on this concept as well. This work has been supported by the Spanish State Research Agency (AEI) and the European Regional Development Fund (ERDF) under Project No. FIS2016-76883-P. 

\appendix

\section*{Appendix}

The following lines are devoted to obtain a power series relating the size of the particle $d$ and the magnitude of the delay $r/c$. This relation allows us to approximate the distance $l$ between the dumbbell's position at time $t$ and at the delayed time $t_{r}$, as a function of the mass center velocity, its derivatives and the particle's size \cite{gri83,gri89}. We begin with the relation
\begin{equation}
d=r \sqrt{1-\left( \frac{l}{r} \right)^2}=r \left(1-\frac{z^2}{2}-\frac{z^4}{8}-...\right),
\label{eq:a1}
\end{equation}
where the variable $z=l/r$ has been introduced. On the other hand, the Eq.~(\ref{eq:6}) can be rewritten as
\begin{equation}
z=\frac{l}{r}=\beta+\frac{a}{2 c^2}r+\frac{\dot{a}}{6 c^3}r^2+\frac{\ddot{a}}{12 c^4}r^3+\frac{\dddot{a}}{120 c^5}r^4...
\label{eq:a2}
\end{equation}
The square of $z$ can then be computed. If we disregard the terms of the third order and higher orders as well, we obtain
\begin{equation}
z^2=\beta^2+\frac{a}{c^2}\beta r+\frac{a^2}{4 c^4}r^2+\frac{\dot{a}}{3 c^3}\beta r^2+O(r^3).
\label{eq:a3}
\end{equation}

Concerning the fourth power of $z$ we can write
\begin{equation}
z^4=\beta^4+\frac{2a}{c^2}\beta^3 r+\frac{3 a^2}{c^4}\beta^2 r^2+\frac{2 \dot{a}}{3c^3}\beta^3 r^2+O(r^3).
\label{eq:a4}
\end{equation}
to the same approximation as before.

Substitution of Eqs.~(\ref{eq:a3}) and (\ref{eq:a4}) into equation (\ref{eq:a1}), after gathering terms, yields
\begin{equation}
d=\left(1-\frac{\beta^2}{2} -\frac{\beta^4}{8} \right)r-\frac{a}{2c^2}\beta \left(1+\frac{\beta^2}{2}\right)r^2-\left(\frac{a^2}{8 c^4}\left(1+\frac{3\beta^2}{2}\right)+\frac{\dot{a} \beta}{6 c^3} \left(1+\frac{\beta^2}{2} \right) \right)r^3+O(r^4).
\label{eq:a5}
\end{equation}

If we consider the non-relativistic limit, by just keeping terms of the first order in $\beta$, we arrive at the approximated relation
\begin{equation}
d=r-\frac{a}{2c^2}\beta r^2-\left(\frac{a^2}{8 c^4}+\frac{\dot{a}}{6 c^3} \beta \right)r^3.
\label{eq:a6}
\end{equation}

\section*{Conflict of Interest}

The authors declare that they have no conflict of interest.

\end{document}